\documentclass[reprint, superscriptaddress, prl, amsmath]{revtex4-2}

\usepackage{graphicx}
\usepackage{dcolumn}
\usepackage{bm}
\usepackage{color}
\usepackage[colorlinks = true, linkcolor=blue, urlcolor  = blue, anchorcolor=blue, citecolor=blue]{hyperref}  \usepackage{siunitx}
\DeclareSIUnit\nounit{{}}
\usepackage[normalem]{ulem}
\usepackage{comment}
\usepackage[utf8]{inputenc}
\usepackage{float}
\usepackage{setspace}
\usepackage{setspace}

\begin{document}

\title{Non-monotonic motion of sliding droplets on strained soft solids}
\author{Youchuang Chao}
\email{ychao@hit.edu.cn}
\altaffiliation{Present address: School of Energy Science and Engineering, Harbin Institute of Technology, Harbin 150001, China}
\affiliation{Max Planck Institute for Dynamics and Self-Organization, G{\"o}ttingen 37077, Germany}
\author{Hansol Jeon}   
\affiliation{Max Planck Institute for Dynamics and Self-Organization, G{\"o}ttingen 37077, Germany}
\author{Stefan Karpitschka}
\email{stefan.karpitschka@uni-konstanz.de}
\affiliation{Max Planck Institute for Dynamics and Self-Organization, G{\"o}ttingen 37077, Germany}
\affiliation{Department of Physics, University of Konstanz, Konstanz 78457, Germany}

\date{\today}

\begin{abstract}
Soft materials are ubiquitous in technological applications that require deformability, for instance, in flexible, water-repellent coatings.
However, the wetting properties of pre-strained soft materials are only beginning to be explored.
Here we study the sliding dynamics of droplets on pre-strained soft silicone gels, both in tension and in compression.
Intriguingly, in compression we find a non-monotonic strain dependence of the sliding speed: mild compressions decelerate the droplets, but stronger compressions lead again to faster droplet motion.
Upon further compression, creases nucleate under the droplets until finally, the entire surface undergoes the creasing instability, causing a ``run-and-stop" motion.
We quantitatively elucidate the speed modification for moderate pre-strains by incremental viscoelasticity, while the acceleration for larger pre-strains turns out to be linked to the solid pressure, presumably through a lubrication effect of expelled oligomers.
\end{abstract}

\maketitle


\textit{Introduction.---}The motion of liquid droplets on soft solids is widely present in nature and everyday life~\cite{style2017elastocapillarity, andreotti2020statics}, e.g., in wetting~\cite{moonen2023versatile} or maceration~\cite{cutting2002maceration} of human skin, impacting also various technological processes ranging from vapor condensation~\cite{sokuler2010softer} to (3D-)printing~\cite{xie2020room} or meniscus formation in atomic force microscopy~\cite{Garcia:SSR2002}.
While droplet motion on rigid substrates is mainly hindered by viscous dissipation inside the liquid phase~\citep{bonn2009wetting, snoeijer2013moving}, on soft materials, viscoelastic dissipation inside the substrate phase may dominate, slowing down the motion by “viscoelastic braking”~\citep{carre1996viscoelastic, long1996static, karpitschka2015droplets,snoeijer2018paradox}. 
Recently, a large number of experimental and theoretical studies have revealed intriguing wetting dynamics on soft surfaces, for instance, stick-slip motion~\citep{pu2008characterization, kajiya2013advancing, park2017self, van2018dynamic, van2020spreading, mokbel2022stick, greve2023stick, jeon2023moving}, droplet durotaxis~\citep{style2013patterning, wang2015patterning, bardall2020gradient, pallares2023stiffness}, 
non-local droplet interaction~\citep{karpitschka2016liquid, pandey2017dynamical}, and capillary-induced phase separation~\cite{cai2021fluid, hauer2023phase, Cai:SM2024}.
Yet, much less is understood about liquid motion on pre-strained materials, which is highly relevant for many applications, e.g., in materials engineering, where soft materials are typically  intended to be stretched~\citep{rogers2010materials} or squeezed~\citep{li2012mechanics}.
Recent work has begun exploring the behavior of liquid droplets on pre-stretched soft solids~\citep{schulman2015liquid, xu2017direct, schulman2018surface, snoeijer2018paradox, xu2018surface, smith-Mannschott2021droplets, nair2023equilibria, kozyreff2023effect, bain2024unravelling}, finding altered wetting ridge and droplet geometries.
Most of those studies investigated static wetting, and much less is known about droplet motion across pre-strained substrates.
Pioneering work has demonstrated that droplets move faster and anisotropically on pre-stretched soft surfaces~\cite{smith-Mannschott2021droplets}.
However, the role of the pre-strain is still not well understood, and, in particular, the dynamics on compressed surfaces remain elusive.

\begin{figure}[b!]
\centering
\includegraphics[width=.45\textwidth]{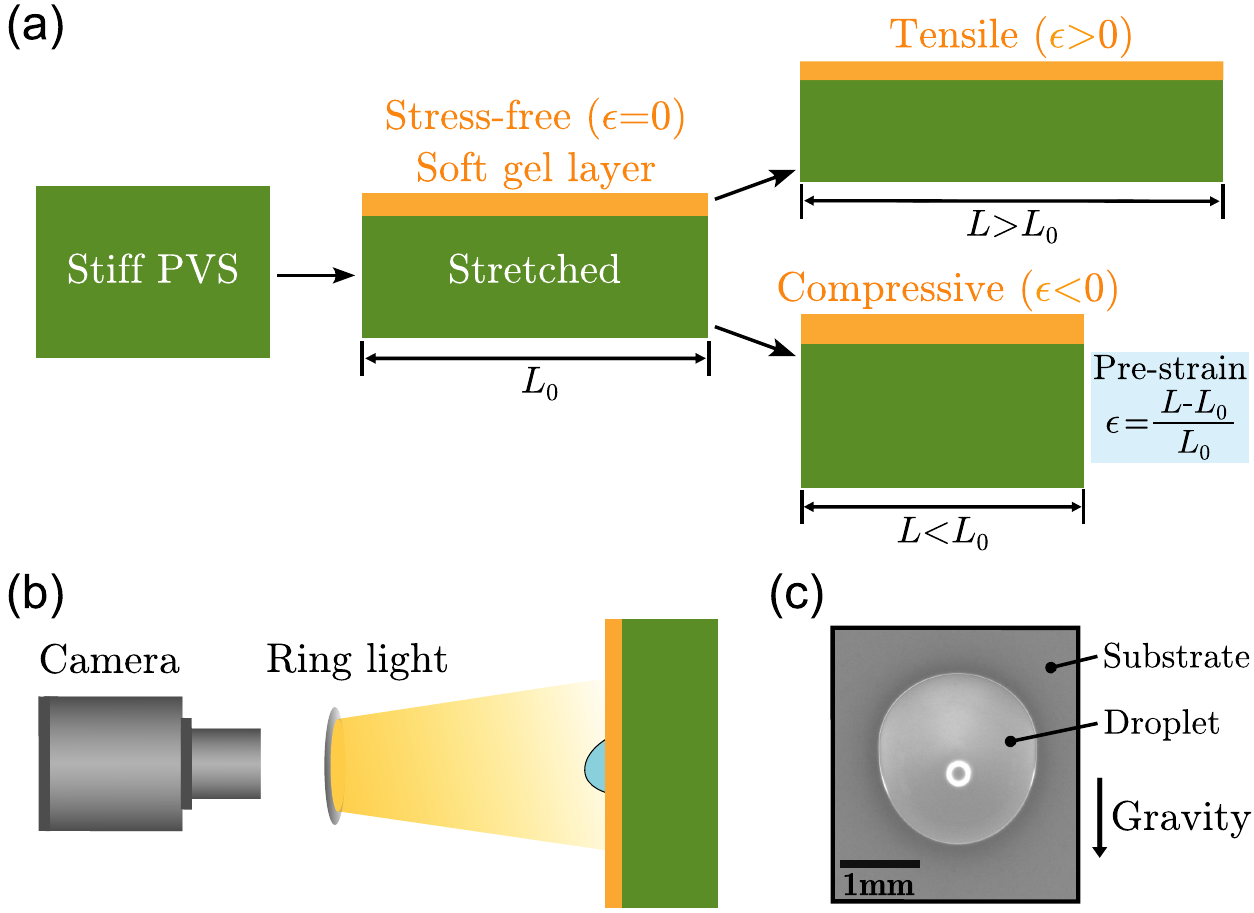}
\caption{(a)~Schematic showing the strategy of uniaxially stretching or compressing soft gels with a defined pre-strain $\epsilon = (L-L_0)/L_0$, where $L_0$ and $L$ are the initial and final lateral lengths, respectively. (b)~Experimental setup: gravity-driven sliding of a droplet on a soft gel, illuminated by a ring light and captured by a digital camera. (c) Typical image of a $\SI{5}{\micro\liter}$ ethylene glycol droplet sliding down a silicone gel. The small white ring in the droplet is a reflection of the ring light.
}
\label{fig:ExperimentalSetup}
\end{figure}

In this Letter, we apply both tensile and compressive pre-strains to soft polymer gels and study how they affect the dynamics of wetting, namely, the motion of ethylene glycol (EG) droplets sliding under the action of gravity (see Fig.~\ref{fig:ExperimentalSetup}).
For tensile pre-strains, we observe a direction-dependent acceleration of droplet motion, similar to previous measurements~\citep{smith-Mannschott2021droplets}.
Unexpectedly, however, for sliding along the direction of a pre-compression, we find a non-monotonic dependence of the speed on the pre-strain: slight compressions decelerate the droplets, as expected from extrapolating the tensile case, while droplets move faster again for larger compressions.
Remarkably, this occurs before the onset of surface instabilities.
Compressing further, droplets are decelerated again and perform a ``run-and-stop" motion, arising from their interaction with periodic creases, i.e., inward folds that emerge on the surfaces of elastic materials beyond a critical compression~\citep{cai2012creasing, hohlfeld2011unfolding, ciarletta2019soft,karpitschka2017cusp,limbeek2021pinning}.
We show that the behavior for moderate pre-strains is quantitatively captured by incremental viscoelasticity~\cite{biot1963incremental}, while the non-linear acceleration for larger compressions is linked to the solid pressure rather than surface stretch, pointing to a lubrication effect from expelled oligomers.

\textit{Experimental methods.---}To prepare pre-strained soft surfaces, we first cure a layer of silicone gel (thickness $h_0 \sim \SI{1.3}{\milli\meter}$) on top of a uniaxially-stretched, stiff polyvinyl siloxane (PVS) rubber sheet (Zhermack Elite Double 22, shear modulus $G_0 \sim \SI{100}{\kilo\pascal}$).
We use two different silicone gels: i) Gelest DMS-V31+HMS-053 (Gelest) with $G_0 \sim \SI{2}{\kilo\pascal}$, and ii) Dow Corning CY52-276A/B (Dow) with $G_0 \sim \SI{1}{\kilo\pascal}$.
The pre-strain $\epsilon$ of the soft gel is adjusted by either additionally stretching ($\epsilon > 0$) or partially releasing ($\epsilon < 0$) the the support with a micrometer stage~\citep{mora2011surface, cai2012creasing, limbeek2021pinning} (Fig.~\ref{fig:ExperimentalSetup}a).
To minimize the influence of long-term trends in material properties, we conduct all measurements at least one week after gel preparation and finish each group of measurements within one day, rationalized by a long-term rheometry experiment (see Supplemental Material~\cite{suppl}).
For each pre-strain, ethylene glycol (EG) droplets with a defined volume are gently deposited onto the silicone gel.
Substrate thickness and droplet radius ($R\sim\mathcal{O}(\SI{1}{mm})$) are much larger than the elastocapillary length $\ell_e \sim \SI{0.037}{\milli\meter}$, estimated by $\gamma\,\sin\theta/G_0$ with $G_0 \sim \SI{1.3} {\kilo\pascal}$, surface tension of EG, $\gamma  = \SI{48}{\milli\newton\per\metre}$~\citep{karpitschka2016liquid}, and an apparent contact angle $\theta\sim \SI{85}{\degree}$.
After droplets reach equilibrium (around 30 min), the setup is rotated so that gravity acts tangentially to the surface.
After a short while, the droplets move stationarily and we record their motion by a digital camera (PointGray Grasshopper2) attached to a telecentric lens (1.0x, working distance 62.2mm, Thorlabs) or a macro lens (Nikon AF Micro Nikkor 60 mm f=2:8D), illuminated by a ring light (Amscope), as shown in Fig.~\ref{fig:ExperimentalSetup}(b) and~1(c). Measurements are performed from tensile to compressive pre-strains, and samples are discarded after creasing to exclude artifacts due to scars~\cite{limbeek2021pinning} (see Supplemental Material~\cite{suppl} for detailed information).

\begin{figure}
\centering
\includegraphics[width=0.48\textwidth]{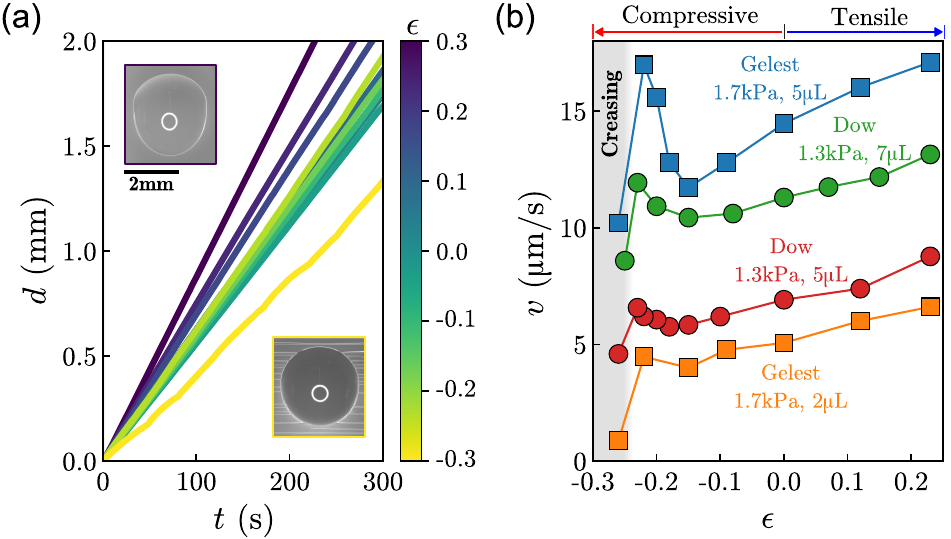}
\caption{Ethylene glycol droplets sliding down silicone gels in the direction parallel to the applied pre-strain $\epsilon$ ($x$-direction).
(a)~Droplet position $d$ as a function of time $t$ for various $\epsilon$ ($\SI{5}{\micro\liter}$ droplets on $\sim\SI{1.3}{\kilo\pascal}$ Dow gel). Insets: droplets for $\epsilon = 0.23$ (top) and $\epsilon = -0.26$ (bottom; creases appear as horizontal lines in the background).
(b)~Non-monotonic dependence of the sliding speed $v$ on $\epsilon$ from $-0.26$ to $0.23$, for different drop sizes ($2$, $5$, $\SI{7}{\micro\liter}$) on two different gels (Gelest, $\sim \SI{1.7}{\kilo\pascal}$; Dow, $\sim \SI{1.3}{\kilo\pascal}$). The grey region indicates where creasing occurs.
}
\label{fig:Sliding1}
\end{figure}

\textit{Non-monotonic sliding behavior.---}
Fig.~\ref{fig:Sliding1}(a) shows the position $d$ of $\SI{5}{\micro\liter}$ ethylene glycol (EG) droplets sliding down a Dow Corning CY52-276 silicone gel with $G_0\sim \SI{1.3}{\kilo\pascal}$, pre-strained along the axis of droplet motion, as a function of time $t$ for various pre-strains $\epsilon$ (color bar).
For mildly pre-strained surfaces, droplets slide with a constant, strain-dependent speed (see Movie S1).
For the largest compression (yellow curve), the droplet moves at a much smaller and non-constant speed, caused by creases: due to a non-linear surface instability~\cite{cai2012creasing, hohlfeld2011unfolding, ciarletta2019soft,karpitschka2017cusp,limbeek2021pinning}, the surface deforms and folds periodically into self-contacts (bright horizontal lines in the lower inset).

Fig.~\ref{fig:Sliding1}(b) shows the mean sliding speed $v$ as a function of pre-strain $\epsilon$, for droplets with different sizes ($2$, $5$, and $\SI{7}{\micro\liter}$), sliding down two different gels (Gelest, $G_0\sim\SI{1.7}{\kilo\pascal}$, and Dow, $G_0\sim\SI{1.3}{\kilo\pascal}$), under a wide range of pre-strain conditions $\epsilon \in [-0.30, 0.30]$.
We begin with the more common, stretched situation ($\epsilon > 0$, right half of Fig.~\ref{fig:Sliding1}b), where we observe a monotonic increase of sliding speed $v$ with $\epsilon$.
This observation is consistent with a recent study~\citep{smith-Mannschott2021droplets}, which reported that droplets move faster in the direction parallel than in the direction perpendicular to the pre-stretch.
In compression ($\epsilon <0$), however, $v$ depends non-monotonically on $\epsilon$.
Initially, the trend from the tensile regime is continued, but after reaching a minimum around $\epsilon \sim -0.18$, the droplet speed increases again.
Notably, this occurs before the onset of creasing (gray gradient region in Fig.~\ref{fig:Sliding1}b).

Beyond a critical compression $\epsilon_c \sim -0.26$, the mean speed suddenly decreases significantly (Fig.~\ref{fig:Sliding1}b, grey gradient region). The irregularity is more pronounced for small droplets, up to a ``run-and-stop" motion ($\SI{2}{\micro\liter}$ and $\SI{1}{\micro\liter}$, Fig.~\ref{fig:S:RunStopMotion} in Supplemental Material and Movie~S2~\cite{suppl}).
Apparently, the formation of creases perpendicular to the direction of the compression~\citep{gent1999surface, hong2009formation} leads to a recurrent pinning-depinning motion of the three-phase contact line. 
For even smaller droplets ($\sim \SI{0.2}{\micro\liter}$ for instance),  the droplets are arrested completely ($\SI{0.2}{\micro\liter}$, Fig.~\ref{fig:S:RunStopMotion} in Supplemental Material~\cite{suppl}).
In this context, we observe that the crease instability is  affected by the presence of droplets: for $-0.23 \gtrsim\epsilon\gtrsim -0.26$, creases nucleate under the sessile droplets, prior to the onset of the global instability (Fig.~\ref{fig:S:CreasingFormation} in Supplemental Material~\cite{suppl}).
Hence, the abrupt drop of the sliding speed for small $\epsilon$ (Fig.~\ref{fig:Sliding1}) is attributed to sub-critical nucleation of creases under the droplets.

\begin{figure}
\centering
\includegraphics[width=0.48\textwidth]{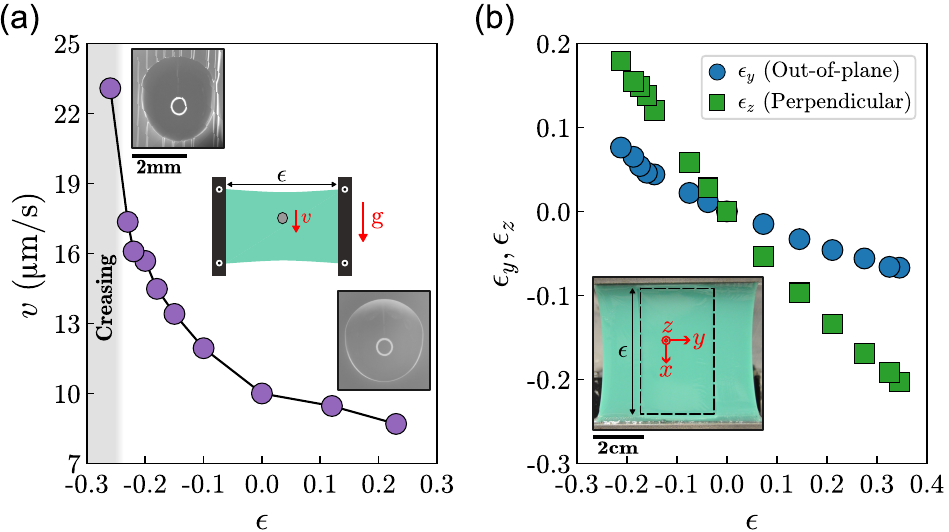}
\caption{Droplets sliding in the direction perpendicular to the applied pre-strain $\epsilon$ ($y$-direction).
(a)~$v$ \textit{vs} $\epsilon$ for $\SI{5}{\micro\liter}$ on a $\sim \SI{1.5}{\kilo\pascal}$ Dow gel. Insets: Images of droplets for $\epsilon =-0.26$ (top, creases appear as vertical lines in the background) and $0.23$ (bottom). 
(b)~Strains in the directions perpendicular to the applied strain, in-plane ($\epsilon_y$, blue circles) and out-of-plane ($\epsilon_z$, green squares) \textit{vs} applied strain ($\epsilon$, $x$-direction).
Inset: Photograph of a sample, indicating the axes.
}
\label{fig:Sliding2}
\end{figure}

\textit{Anisotropic droplet motion.---}In the light of the previously observed anisotropy of the droplet mobility on stretched surfaces~\cite{smith-Mannschott2021droplets}, we next examine the sliding of $\SI{5}{\micro\liter}$ EG droplets on the Dow gel ($\sim \SI{1.5}{\kilo\pascal}$) perpendicular to the direction of the applied pre-strain, i.e., in $y$ direction, as shown in Fig.~\ref{fig:Sliding2}(a).
Note that the slightly higher modulus of this fresh sample is insignificant as compared to the impact of pre-straining.
In this case, we observe a monotonic dependence of the sliding speed $v$ on the pre-strain $\epsilon$ that is opposite to the parallel case: now, stretching decelerates droplets, whereas compression accelerates them.
Intriguingly, that trend continues as creases start to form: oriented parallel to the direction of motion, creases are not impeding the droplet motion.

In all our experiments, we apply a far-field uniaxial pre-strain at two opposite edges of the PVS support rubber.
Accordingly, the gel layer around the center of the support, where drop sliding is measured, is pre-strained biaxially, where we define the direction of the applied strain $\epsilon$ as the $x$-direction.
Fig.~\ref{fig:Sliding2}(b) shows $\epsilon_y$, the strain in $y$ direction, measured from the displacement of marks on the support layer, as a function of the applied strain ($x$-direction).
The vertical ($z$-)strain $\epsilon_z$ is calculated from the $x$- and $y$-strains by assuming an incompressible silicone gel, in close agreement with the reported Poisson ratio $\nu\sim 0.48$~\cite{xu2017direct}.
Recent experiments have demonstrated that wetting ridge sections oriented perpendicular to the droplet motion dominate viscoelastic braking~\cite{smith-Mannschott2021droplets, xue2024droplets}.
Thus, if viscoelastic braking according to~\cite{smith-Mannschott2021droplets} determines the speed, the strain in $y$ direction ($\epsilon_{||}$) should dominate.
While the observed dependency is qualitatively consistent with Ref.~\cite{smith-Mannschott2021droplets} and Fig.~\ref{fig:Sliding1}(b) for tensile and small compressive strains, however, plotting speed $v$ \textit{vs} $\epsilon_{||}$  cannot collapse the data:
The increase of $v$ for negative $\epsilon$ is much larger than for comparable stretching in the parallel pre-strain case (Fig.~\ref{fig:S:SpeedVSparallelStrain} in Supplemental Material~\cite{suppl}), so a dependency on $\epsilon_{||}$ alone cannot explain the observed speeds.

\begin{figure}
\centering
\includegraphics[width=0.48\textwidth]{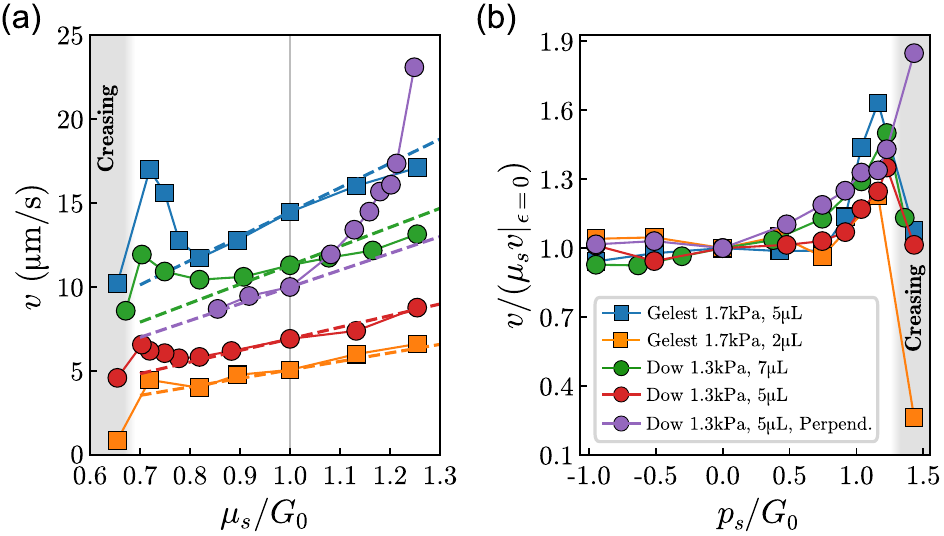}
\caption{(a) Sliding speed $v$ \textit{vs} effective surface modulus $\mu_s$ over shear modulus $G_0$. Creases appear well before $\mu_s\rightarrow 0$, indicated by the gray region. Linear visco-elasto-capillary theory predicts a linear relation between $v$ and $\mu_s/G_0$ (dashed lines). 
(b) Sliding speed $v$ normalized by $v|_{\epsilon=0}$ (the speed at $\epsilon=0$) and the effective surface modulus $\mu_s$ as a function of the solid pressure $p_s$ over $G_0$. Sliding speeds are well captured by incremental linear viscoelasticity (i.e., $v/(\mu_s v|_{\epsilon=0})\sim 1$) for negative or slightly positive pressure, but increase strongly for large positive pressure, irrespective of the surface strain parallel to the direction of sliding (see Fig.~S3 in Supplemental Material~\cite{suppl}).
}
\label{fig:vvsmus}
\end{figure}

\emph{Discussion.---}The non-monotonic speed-strain dependence is unexpected, since it is incompatible with the classical, linear viscoelastic, soft wetting theory~\citep{long1996static, karpitschka2015droplets}.
To examine this discrepancy, it is instructive to first analyze the response of a pre-strained material to the capillary pinching of the droplet by linearizing the equations of elasticity around a state of homogeneous finite pre-strain~\cite{biot1959folding,biot1963incremental,biot1963surface}. Namely, this incremental response is analogous to the linear response of the reference configuration, but with an effective modulus $\mu$ and an additional term in the equilibrium condition that depend on the pre-strain~\cite{biot1959folding,biot1963incremental,biot1963surface} (see Eq.~S1 of Supplemental Material~\cite{suppl}). In the case of an incompressible half-space ($\nu = 1/2$), the surface response to localized line tractions changes only in amplitude, quantified by an effective surface modulus $\mu_s$, expressed as
\begin{align}
	\mu_s &= \mu\,\left[(1-\zeta)^{1/2}(1+\zeta)^{3/2}-1\right]/\zeta, \\
		\mu   &= \frac{G_0}{2}\left(\lambda_1^2 + \lambda_2^2\right),
\end{align}
where $\lambda_1$ and $\lambda_2$ are two principle elongations, aligned with the horizontal (surface-parallel, $x$- or $y$-direction) and vertical axis ($z$-direction) of the incremental deformation.
In our configuration, $\lambda_1\approx 1+\epsilon$ or $\lambda_1\approx (1+\epsilon)^{-0.25}$, for sliding parallel or perpendicular to the pre-strain $\epsilon$, respectively, and $\lambda_2\approx (1+\epsilon)^{-0.75}$ for both cases.
$\zeta \approx (\lambda_2^2 - \lambda_1^2)/(\lambda_2^2 + \lambda_1^2)$ is the effective pre-compression of the material~\cite{biot1963surface}.
For sliding parallel to the pre-strain, $\mu(\epsilon)$ is non-monotonic, increasing by $\sim 15\%$ toward the edges of the investigated range of $\epsilon$, while $\mu_s (\epsilon)$ depends monotonically on $\epsilon$, vanishing at a critical compression~\cite{biot1963surface} where the surface becomes linearly unstable (see Fig.~\ref{fig:S:EffectiveModulusVSStrain} in Supplemental Material~\cite{suppl}).
However, this linear instability is preceded by aforementioned non-linear creasing instability, in which the surface folds inward, forming a self-contact~\cite{ciarletta2019soft,karpitschka2017cusp,limbeek2021pinning,dervaux2012mechanical, cai2012creasing,tallinen2013surface} (gray zones in Figs.~\ref{fig:Sliding1}b,~\ref{fig:Sliding2}a and inset images).
Since dissipation scales inversely to the height of the wetting ridge~\cite{karpitschka2015droplets, khattak2022direct} that is governed by the elastocapillary length $\ell_e = \gamma\,\sin\theta / \mu_s$~\cite{long1996static,style2013universal}, incremental elasticity predicts that $v\sim\mu_s$, so we plot $v$ as a function of $\mu_s/G_0$ in Fig.~\ref{fig:vvsmus}(a).
For small $|\epsilon|$, where $\mu_s/G_0$ remains close to unity, the droplet sliding speed indeed follows the expected trend $v\sim\mu_s$ (dashed lines in Fig.~\ref{fig:vvsmus}a).
Only the non-linear increase of $v$ toward large $|\epsilon|$ ($\mu_s/G_0\not\approx 1$), which, importantly, occurs at opposite $\mu_s$ for parallel and perpendicular sliding directions, is apparently not captured by incremental elasticity alone.

This is neither a consequence of the drop aspect ratio, which changes only very mildly (see Fig.~\ref{fig:S:DropAspectRatio} in Supplemental Material~\cite{suppl}), nor a consequence of the finite thickness of the soft layer:
In our experiments, the largest elastocapillary length $\ell_e\sim \SI{0.037}{\milli\meter}$ (Dow, $\sim \SI{1.3}{\kilo\pascal}$) remains much smaller than the minimum gel thickness $h_{0, min}\sim \SI{1.12}{\milli\meter}$ ($\epsilon = 0.23$).
Therefore, energy dissipation dominates near the ridge tip~\citep{van2020spreading,zhao2018geometrical, khattak2022direct,essink2024wetting}, implying an effective elastic half-space response.
Including the effect of finite thickness in incremental elasticity is straight forward, but it has only a minor influence on the ridge height (see Eqs.~\ref{eq:S:K22},~\ref{eq:S:profile} and Fig.~\ref{fig:S:heights} in Supplemental Material~\cite{suppl}).

The observation of a non-monotonic or strongly increasing speed for large pre-compression is also incompatible with the explanation in terms of a strain-dependent solid surface tension~\citep{smith-Mannschott2021droplets}. In particular, Smith-Mannschott \textit{et al.}~\citep{smith-Mannschott2021droplets} showed that, when the silicone gel is pre-stretched, the height of the wetting ridge is smaller along the direction of the pre-stretch as compared to the opposite direction, and attributed this effect to a strain-dependent solid surface tension.
In that case, the speed should mainly depend on $\epsilon_{||}$, independent of the direction of the applied strain, which is not the case, as shown in Fig.~\ref{fig:S:SpeedVSparallelStrain} in Supplemental Material~\cite{suppl}.
Extrapolating to negative pre-strains, one would expect a monotonic decrease in sliding speeds unless the surface constitutional relation exhibits a non-monotonic dependence of surface tension on strain.
The logarithmic term suggested in Ref.~\citep{pandey2020singular} would offer a non-monotonicity in compression, but currently there is neither a microscopic model that would motivate such a particular dependency, nor a strain-dependent surface tension of soft polymer gels in general~\cite{schulman2018surface}.
Measurements are restricted to singular cases like contact lines in wetting or adhesion~\cite{xu2017direct,snoeijer2018paradox,schulman2018surface,xu2018surface, heyden2021contact,smith-Mannschott2021droplets}, inherently impeded by a logarithmic or algebraic singularity for statics or dynamics, respectively; Thus, faithful resolution of solid angles requires spatial resolutions of at least $\mathcal{O}(\ell_e/20)\lesssim\SI{1}{\micro\meter}$~\cite{pandey2020singular, jeon2023moving}.
Fig.~\ref{fig:S:profiles} in Supplemental Material illustrates that this pertains to the impact of the pre-stretch. 
The present results, where majority of the data is readily explained by incremental linear viscoelasticity~\cite{biot1963surface} (Fig.~\ref{fig:vvsmus}a), render the strain-dependent solid surface tension more unlikely.

Notably, recent experimental and theoretical work has demonstrated the importance of poroelastic effects in static and dynamic wetting of soft polymer gels~\cite{hourlier2017role,zhao2018growth,xu2020viscoelastic,cai2021fluid, cai2022swelling,hauer2023phase,flapper2023reversal,xue2024droplets,Cai:SM2024}.
Soft silicone gels typically contain a significant portion ($\sim 50\%$) of mobile chains inside the crosslinked network, i.e., ``extractables''~\citep{nandi2005swelling, hourlier2017role, glover2020extracting, cai2021fluid}.
These extractables may phase-separate into a liquid skirt around the wetting ridge~\citep{jensen2015wetting, cai2021fluid, hauer2023phase} or migrate to contacting materials~\citep{cai2022swelling}. Indeed, we observe a significant long-term trend in the rheometry when the gel is in contact with the PVS support (see Figs.~\ref{fig:S:ExtractableTest} and \ref{fig:S:RheologicalTest} in Supplemental Material~\citep{suppl}), and therefore perform the experiments only after $\sim 6$~days of sample aging when rheological properties became stationary.
An oil skirt, whose size depends on the speed of moving contact line and the degree of pre-swelling, can reduce the viscoelastic dissipation in the solid~\cite{hauer2023phase}.

Importantly, the equilibrium chemical potential in the gel depends on the isotropic part of the elastic stress tensor~\cite{flory1942thermodynamics,flory1943statistical,flapper2023reversal}, i.e., the solid pressure $p_s$, which is dominated by the pre-strain $\epsilon$. For a neo-Hookean model,
\begin{equation}
p_s \sim - G_0 [(1+\epsilon)^2 + (1+\epsilon)^{-0.5} - 2(1+\epsilon)^{-1.5}],
\end{equation}
where we employed the no-stress boundary condition $S_{22}=0$ at the free surface~\cite{biot1963surface}.
Thus, the pre-strain in our samples is expected to alter the equilibrium between the dissolved liquid and the skirt.
To support this hypothesis, we normalize the droplet sliding speed $v$ with the speed without pre-strain, $v|_{\epsilon=0}$, and remove the expected speed dependence from incremental elasticity by scaling with $\mu_s$.
Plotting the scaled speed $v/(\mu_s v|_{\epsilon=0})$ over the solid pressure $p_s/G_0$ (Fig.~\ref{fig:vvsmus}b), the non-linear acceleration is indeed observed for strongly positive $p_s/G_0$, where the chemical potential difference would drive more liquid into the skirt, similar to squeezing liquids out of a sponge.
The experimental data is collapsed to a significant degree.
A qualitative confirmation of our interpretation is obtained from a test experiment immediately after sample preparation, where extractables are expected to be more abundant in the gel. There, we observe significantly larger sliding speeds and a shift of the speed minimum toward positive pre-strains (Fig.~\ref{fig:S:freshsample} in Supplemental Material~\cite{suppl}).
This cannot be explained by the slightly different rheological parameters, but is consistent with a larger amount of swelling fluid in the sample.
Thus, developing a quantitative model for the visco-poro-elasto-capillary dynamics with phase separation into ``four-phase contact zones''~\cite{jensen2015wetting} will be an important step for future research.

\textit{Conclusion.---}In summary, we have studied how a tensile or a compressive pre-strain affects the elasto-wetting dynamics of droplets sliding down soft silicone gels.
For moderate pre-strains, the sliding speeds can be explained quantitatively by incremental elasticity, assuming a strain-independent solid surface tension.
Unexpectedly, we find a non-monotonic regime for droplet motion along large imposed compressive pre-strains, and a substantial non-linear acceleration for sliding perpendicular to the applied strain.
These phenomena occur before the onset of surface instabilities and can not be rationalized by employing the theories of classical visco-elasto-capillary wetting~\citep{long1996static, karpitschka2015droplets}, even with incremental strains or strain-dependent surface tension~\citep{smith-Mannschott2021droplets}. 
The origin of the large-strain acceleration is likely due to a lubrication effect resulting from poroelastic phase separation, as it aligns with the solid pressure and thus the chemical potential in the solid, rather than the surface strain along the sliding direction.
Scrutinizing further the role of large compressive pre-strains on soft wetting dynamics will require
thermodynamically consistent models that resolve those phases in arbitrary topology.
Our findings, however, elucidate the importance of both solid pressures and surface instabilities for controlling droplet motion by pre-strains, enriching the current understanding of elasto-capillarity-related fluid-structure interaction applications where new soft materials are commonly tested and operated under pre-strained and/or wet conditions~\citep{sun2012highly, li2017tough}.

\begin{acknowledgments}
Y.C. acknowledges support through an Alexander von Humboldt Fellowship. S.K. and H.J. acknowledge funding from the German research foundation (DFG, Project No. KA4747/2-1). S.K. acknowledges helpful discussions with J. H. Snoeijer and A. Pandey. We also would like to thank K. Hantke and W. Keiderling for assistance with the experimental setup.
\end{acknowledgments}

\end{document}